\documentclass[utf8]{frontiersFPHY} 

\usepackage{url,hyperref,lineno,microtype}
\usepackage[onehalfspacing]{setspace}
\usepackage{aas}
\definecolor{cadmiumred}{rgb}{0.89, 0.0, 0.13}
\definecolor{bgreen}{rgb}{0.0, 0.26, 0.15}
\definecolor{scc}{rgb}{0.0, 0.26, 0.15}
\definecolor{alex}{rgb}{0.0, 0.28, 0.67}
\def\keyFont{\fontsize{8}{11}\helveticabold }
\def\firstAuthorLast{Brescia {et~al.}} 
\def\Authors{Massimo Brescia\,$^{1*}$, Stefano Cavuoti\,$^{1,2*}$, Oleksandra Razim\,$^2$, Valeria Amaro\,$^2$,  Giuseppe Riccio\,$^1$, Giuseppe Longo\,$^2$}
\def\Address{$^1$INAF Astronomical Observatory of Capodimonte, Napoli, Italy\\
$^2$Department of Physics Ettore Pancini, University Federico II, Napoli, Italy}

\def\corrAuthor{Massimo Brescia\\ INAF Astronomical Observatory of Capodimonte, Salita Moiariello 16, Napoli, I-80131, Italy}
\def\corrEmail{massimo.brescia@inaf.it}

\begin{document}
\onecolumn
\firstpage{1}

\title[Photometric redshifts with machine learning]{Photometric redshifts with machine learning, lights and shadows on a complex data science use case}
\author[\firstAuthorLast ]{\Authors} 
\address{} 
\correspondance{} 

\extraAuth{ Stefano Cavuoti \\ INAF Astronomical Observatory of Capodimonte, Salita Moiariello 16,  Napoli, I-80131, Italy\\  stefano.cavuoti@inaf.it}

\maketitle

\begin{abstract}
The current role of data-driven science is constantly increasing its importance within Astrophysics, due to the huge amount of multi-wavelength data collected every day, characterized by complex and high-volume information requiring efficient and as much as possible automated exploration tools.
Furthermore, to accomplish main and legacy science objectives of future or incoming large and deep survey projects, such as JWST, LSST and Euclid, a crucial role is played by an accurate estimation of photometric redshifts, whose knowledge would permit the detection and analysis of extended and peculiar sources by disentangling low-z from high-z sources and would contribute to solve the modern cosmological discrepancies. The recent photometric redshift data challenges, organized within several survey projects, like LSST and Euclid, pushed the exploitation of multi-wavelength and multi-dimensional data observed or ad hoc simulated to improve and optimize the photometric redshifts prediction and statistical characterization based on both SED template fitting and machine learning methodologies. But they also provided a new impetus in the investigation on hybrid and deep learning techniques, aimed at conjugating the positive peculiarities of different methodologies, thus optimizing the estimation accuracy and maximizing the photometric range coverage, particularly important in the high-z regime, where the spectroscopic ground truth is poorly available. In such a context we summarize what learned and proposed in more than a decade of research.

\section{}

\tiny
 \keyFont{ \section{Keywords:} photometric redshifts, machine learning, astroinformatics, data analysis, galaxies} 
\end{abstract}

\section{Introduction}\label{sec:Intro}

Most open questions in cosmology, such as galaxy formation and evolution, the distribution of dark matter, or the understanding of large-scale structure, rely on an accurate estimate of galaxy distances. In the past, such distances could be obtained only for small samples of objects via the displacement (redshift) of spectral features caused by the cosmological expansion, but the time consuming and expensive spectroscopy could not be effectively used either on very faint sources or on large samples of galaxies.
This led to the development of alternative techniques, collectively called \textit{photometric redshift estimation methods}, first proposed by \citet{Baum1962} and better formalised by \citet{Butchins1981} and in the seminal paper of \citet{Connolly1995}.\\
The true turning point, however, came with the era of Sloan Digital Sky Survey (SDSS, \cite{York2000}), the first extensive multi-band and spectroscopic native digital survey of the sky. Among many other applications, this survey made the widest astrophysical community able to explore different approaches to the evaluation of galaxy distances. With its enormous success, SDSS also paved the way to present and future survey projects such as, the Dark Energy Survey (DES, \cite{DES2005}), the Kilo-Degree Survey (KiDS, \cite{deJong2013}), Hyper Suprime-Cam Survey (HSC, \cite{Aihara2018}), Vera C. Rubin Observatory Legacy Survey of Space and Time (LSST, \cite{Abell2009}), Euclid \cite{Laureijs2011}, Cosmic Evolution Survey (COSMOS, \cite{Scoville2007}), James Webb Space Telescope (JWST, \cite{Kauffmann2020}) and Roman Space Telescope \cite{Green2012}, all driven by a new reliance on the possibility to pursue precision cosmology by combining high precision and deep photometry for very large samples of galaxies with a largely incomplete spectroscopic knowledge.\\ 
The field of photometric redshifts (photo-z) estimation benefited from this new wealth of data. The idea behind this line of research is simple: due to the cosmological expansion, the spectrum of a galaxy is stretched towards the red end of the spectrum and, therefore, in a given photometric system, the spectrum of identical galaxies at different distances is weighted differently. In other words: two identical galaxies at different redshifts will have different photometric signatures (magnitudes and colours). 
In practice, things are not so easy, since the function mapping a given photometric space into the redshift space $z$ is complex, depending on many factors (such as morphological type, large scale structure and evolutionary stage) and cannot be uncovered analytically.\\ In the first approximation, photometric redshifts estimation methods can be grouped in two broad branches:
\begin{itemize}
\item[-] \emph{SED template fitting methods}. The redshift is derived by fitting the observed photometry of a galaxy to a set of templates which can be either observed or derived by averaging the spectra of similar galaxies, or computed via synthetic spectroscopy.
\item[-] \emph{Empirical Methods}. This category is characterized by machine learning or data-driven methods that learn how to map photometric space onto $z$ using, in the case of supervised learning, a-priori knowledge provided by a sub-sample of objects for which accurate spectroscopic information is available. Or, alternatively, proceeding to self-organize the photometric information, identifying regions of the parameter space characterized by similarity factors.
\end{itemize}
Already in the early $2000$ it became apparent that machine learning based methods were an ideal and promising tool to deal with this kind of problems \cite{tagliaferri2003,Firth2003}.\\
Over the years, the positive aspects and intrinsic limitations, as well as the complementary nature of both methodological branches resulted evident, depending on a variety of factors \cite{Salvato2019}. For instance, the coverage and sampling of the \emph{observable parameter space}, i.e. an N-dimensional space, where each dimension is defined by an observed photometric quantity (either fluxes, magnitudes or derived colours), the quality of spectroscopic templates, the fraction of peculiar objects, the redshift range, the depth and variety of the photometric information, etc.\\
In what follows the discussion is mostly centered on empirical methods, based on machine learning, by focusing the attention on some aspects of the photometric redshifts estimation problem, which in our opinion seem to be the most relevant.\\
In the following all the quantities related to the photometric redshift error measurement ($\Delta z = z_{phot}-z_{spec}$) are considered as normalized to $(1+z_{spec})$.\\
This work does not claim to be a review on the subject, but rather a synthesis aimed at focusing attention on particular aspects related to the photo-z problem and to the approach based on data-driven methods, which we faced over the years. Aspects that are primarily related to some crucial problems, still open, by highlighting the state of the art of the proposed solutions, both in terms of benefits and critical points. Therefore, this work is primarily aimed at astrophysicists, already familiar with data science techniques (typically astroinformaticians), interested in the problem of prediction and estimation of photometric redshifts with machine learning methodologies. Of course, with the ultimate aim of improving the quality of the photo-z estimation in view of their better scientific exploitation in large astronomical survey projects.

\noindent {\bf Outline:} Sec.~\ref{sec:Leverage} illustrates how machine learning is involved in photo-z estimations and its relevant aspects in this field. In Sec.~\ref{sec:PS} the critical role of the parameter space and the selection of the features is discussed, while in Sec.~\ref{sec:conclusions} we draw some conclusions, projected on next future perspectives.\\

\section{The machine learning leverage on photo-z estimation}\label{sec:Leverage}

In order to be useful, photometric redshifts need to meet strict requirements dictated by the specific application in mind. For example, in the tomographic photo-z bins the estimation error of the true average redshift is required to be less than $\sim0.002$, with a very low outliers rate, in order to be suitable for the cosmic shear estimations \cite{Knox2006,Pasquet2019}. For the LSST survey project, a series of scientific requirements is envisaged, aimed at avoiding that any systematics, in the estimation of the photometric redshifts of several billion galaxies, cannot dominate the statistical background noise of the cosmological sample. In this respect, the requirements specify that the photometric redshift of any individual galaxy should have a bias below 0.003, an estimation error $\sigma_z < 0.02$ and a $3\sigma$ outlier rate below $10\%$ \cite{Schmidt2020}.\\
Furthermore, in the case of gravitational lensing, i.e. the image distortion of background galaxies due to the differential deflection of their light, caused by the masses of foreground sources, the distortion of the coherent shape of galaxies is called the shear of weak lensing and is usually much smaller than the intrinsic ellipticity of galaxies. The measurement of these effects is feasible only in statistical terms, by evaluating the average over a large sample of galaxies, but it is considered one of the most perspective tools to probe the distribution of the dark matter \cite{Mandelbaum2018}. Being less sensitive to the precision of the photometric redshift of individual galaxies,  the three metrics usually adopted to quantify the accuracy of photometric redshifts, i.e. mean bias, scatter, and catastrophic outliers rate are not sufficient to quantify the efficacy of a photo-z method for lensing. For example, some results show that for higher photo-z's the calibration bias in the case of galaxy-galaxy lensing can be as high as $30\%$, although the average redshift bias is well below the dispersion \cite{Mandelbaum2008}. The main reason is the non-linear dependence of the surface density on the redshift of the source, which induces an asymmetrical increase of the photo-z estimation errors. Therefore,	the	error	associated	with	photometric	redshift	measurements is a function of the type and apparent magnitude of the galaxy, with the lensing calibration being very sensitive to the details of the uncertainty distribution on the photo-z estimation \cite{Fu2018,Mandelbaum2008,Ma2006}.\\
These are just few examples of the crucial role played by photometric redshifts in Astrophysics, which justifies the constant and massive proliferation of proposed solutions to optimize their accurate and reliable estimation.\\
The strong dependence of lensing accuracy and galaxy characterization from photo-z precision, together with the required availability of a wide sample of sources, pushed many large survey projects, like LSST, KiDS and Euclid to perform an extensive investigation campaign dedicated to a comparison among all the most popular photo-z methods \cite{Schmidt2020,Hildebrandt2017,EuclidChallenge}.\\

\subsection{General aspects of the photo-z estimation with machine earning}\label{sec:Plethora}

In order to face the relationship between photo-z and machine learning, we start by introducing a series of general aspects.\\ 
Photo-z estimation has now become an indispensable tool in extragalactic astronomy, as the pace of galaxy detection in imaging surveys far outstrips the rate at which follow-up spectroscopy can be performed. A wide plethora of methods and techniques have been and are studied and experimented on a large variety of all-sky multi-band surveys, based either on physical template models fitting the Spectral Energy Distributions (SED) or on empirical explorations of the photometric parameter space, trying to learn its hidden cross-correlation with spectroscopic redshifts, provided for a limited sample of objects. In general, the machine learning (ML) based techniques are able to produce a high-quality photo-z estimation within the photometric ranges imposed by the spectroscopic training set, but less capable to reach the same photo-z estimation quality outside those ranges. Nevertheless, the positive contribution of data-driven methodologies to the estimation of distances for galaxies and peculiar objects, such as quasars \citep{Baron2019,Fluke2020}, is well known. Without claiming to be exhaustive, we can cite the following methods proposed in the literature, which testify to their diversity of approach:
\begin{itemize}
    \item  supervised feed-forward neural networks \cite{Collister2004,Vanzella2004,Brescia2013,Brescia2014,Brescia2015,Brescia2019,Cavuoti2014,Sadeh2016,Almosallam2016};
    \item self-adaptive methods for the detection and removal of anomalies from photometric and spectroscopic data \cite{Hoyle2015,Baron2017,Reis2019};
    \item Support Vector Machines \citep{Zheng2012,Zhang2014,Han2016,Jones2017};
    \item tree-based \cite{CarrascoKind2013,Jouvel2017,Meshcheryakov2018};
    \item k-Nearest Neighbours \cite{Graham2018,Curran2020};
    \item Gaussian processes \cite{Almosallam2016,Bonfield2010}; 
    \item Mixture Density Networks \cite{Ansari2020};
    \item unsupervised models for clustering and for estimating the coverage of the parameter space \cite{Way2012,Masters2015,Stensbo-Smidt2017} or for calibration purposes \cite{Hildebrandt2010,Masters2015,Wright2020};
    \item deep Neural Networks, especially relevant for the photo-z prediction from images \cite{DIsanto2018b,Pasquet2019,Chong2019};
    \item hybrid methods for the selection of photometric redshifts considered particularly accurate and useful for cosmological purposes \cite{Leistedt2017,Bonnett2016,Morrison2017,Fu2018,Salvato2019}.
    \end{itemize}

For the sake of completeness, also techniques based on physical priors knowledge in the form of template spectral energy distributions, the so-called SED template fitting methods, able to adapt to the observed flows and to extrapolate the redshift through chi-square minimization \cite{Arnouts1999,Bolzonella2000,Brammer2008}, are available in an equally rich variety of nuances, as well as several hybrid methods exploiting the Bayesian inference and nested sampling techniques \cite{Benitez2000,Goodman2010,Feroz2019}.\\
The crucial aspect of supervised machine learning methods applied to photo-z prediction is that they require a knowledge base to learn the complex relationship between broad-band photometry and distance, mainly composed by a spectroscopic redshift counterpart sub-sample of the photometric sources used for training, validation and blind testing. When it is available a sufficient spectroscopic coverage of the photometric parameter space, the ML models demonstrated a high photo-z prediction accuracy, although within the limits imposed by the spectroscopic sample \cite{Brescia2019,Schmidt2020,EuclidChallenge}.\\ 
A weakness of these  methods is that most ML models result often biased in presence of large numbers of missing data within the training set.
This can be easily understood by realizing that these models require always the definition of a metric distance that, in order to work properly, needs geometrical varieties characterized by the same dimensionality. 
In Astronomy the problem is further complicated by the fact that missing data can be of different types: truly missing data (e.g. a given object has not been observed in one or more bands) or upper limits (i.e. the object has been observed but not detected) and therefore results as a Not-a-Number in the dataset. Obviously this second type of data carries information on the properties of the objects, which need to be taken into account wherever possible.
In most cases, when only a relatively low fraction of the data is plagued by missing values, it is an acceptable compromise to reject the incomplete data or to apply any imputation technique \cite{Ejaz2020}. This approach, however, is not viable in all those cases where a high amount of data is incomplete. In these cases a more reliable solution is to use methods which are less sensitive to the problem such as, for instance, the Probabilistic Random Forest \cite{Reis2019}.\\
Finally, there are no standard rules for the random splitting of the knowledge base in training, validation and testing subsets, neither in terms of relative percentages nor for the extraction mechanism (random extraction, decimation, etc). 
The optimal partition and sampling strategy can be pursued on a trial and error base, but, as a rule of thumb, in presence of a congruous data amount (at least few thousands), relative percentages of, respectively, $60\%$, $20\%$ and $20\%$, randomly extracted, are a standard choice.\\

\subsection{Unity is strength. Virtuous synergies among methodologies}\label{sec:Synergies}
As known, the SED template fitting methods, based on the adaptation of multi-wavelength photometric observations of objects to a synthetic or observed model SED library, are able to simultaneously provide the estimate of photometric redshifts, the probability density function and the spectral type of each source. However, such methods suffer in particular from the potential mismatch among the synthetic models used for the fitting and the physical properties of the selected sample of observed galaxies \citep{Abdalla2011}, from colour/redshift degeneracy, bias induced by the attenuation law \cite{Calzetti2012,Calzetti2015} and from the incompleteness of the model template library available. Nonetheless, they have the prerogative of being able to derive an estimate of the redshifts, theoretically without any limit in photometric depth.\\
Conversely, the supervised ML methods suffer from the difficulty of obtaining good performances outside the regions of the observed parameter space, adequately covered by the reference spectroscopic sample. On the other hand, it has been amply demonstrated that, where a sufficiently adequate knowledge base is available, most ML methods are more accurate than SED fitting methods in terms of redshift prediction \citep{Hildebrandt2010,Cavuoti2012,Brescia2019}.
This basic complementarity between the two methodologies has recently inspired the hybridization of methods for estimating photometric redshifts, in which it was possible to combine the positive aspects of both techniques in order to overcome their intrinsic limits. For example, the CPz method \cite{Fotopoulou2018} combines the two techniques to derive an automatic method to identify different types of sources, estimating their photometric redshifts and identifying anomalies.
In another case, a hierarchical Bayesian combination of redshift estimates from different models results capable of producing a more accurate estimate of the performance of individual models \cite{Duncan2018}.\\
\citet{Cavuoti2017} started from the assumption that the spectral type classification provided by the SED fitting method allows to derive statistical errors as a function of the spectral type also for ML models, thus making possible a more accurate and specific characterization of the prediction errors. In other words, it is possible to assign a specific spectral class to each source and build specialized (i.e. gated expert) regression models for each spectral type class, thus refining the photometric redshift estimation process. At the end of the hybridization process and the improvement of the quality of the redshifts obtained by the expert ML regression estimator on single spectral types of objects, the proposed method was able to reduce the overall photo-z estimation error by more than $10\%$, compared to the whole blind test set. This improvement was mainly a consequence of the reduction in the percentage of outliers. This result, combined with the prerogative of total arbitrariness in the choice of SED fitting and ML methods to be used, demonstrates the potential of the idea of optimizing the accuracy of the photometric redshifts estimation through the mutual cooperation among theoretical and empirical methods \cite{Cavuoti2017}.\\
More recently, \citet{Soo2021} analyzed and optimized the hybrid empirical-template method Delight \cite{Leistedt2017} on a subset of the early PAUS (Physics of the Accelerating Universe Survey) data release \cite{Eriksen2019}. Delight is an algorithm for the determination of photo-z that combines template-based and machine learning techniques. Delight constructs a large collection of SED templates from training data, with a template SED library as a learning guide for the model. \citet{Soo2021} optimized Delight by calibrating its $40$ narrow bands with six broad bands in the COSMOS field and by performing an interesting analysis of outliers, obtaining, as preliminary result, that narrow band filters produce a large amount of outliers \cite{Soo2021}.
A fact that was experimented by performing photo-z predictions with machine learning models with the $30$-band COSMOS data, composed by a variety of broad and narrow bands, and which was also recently confirmed by \citet{Razim2021}.\\

\subsection{Combined predictions of photo-z and galaxy properties}\label{sec:Combined}
In the context of upcoming and future large survey projects, such as LSST, Euclid and JWST, which will extend our knowledge of the dependence of galaxy populations on environments, as well as the characterization of large scale structures, the determination of star-forming activity, such as the Star Formation rate (SFR) and stellar mass from UV, optical or IR luminosity, will be crucial. Their traditional study was based on complex models and priors on the properties of the galaxy, which limited their capability to accurately describe peculiar categories of extended sources, such as passive galaxies, which are of particular interest in the study of large-scale structures. It is worth emphasizing that the derivation of such physical quantities of galaxies cannot be separated from an accurate and reliable estimate of their distances, and that the redshifts and the physical properties of galaxies are intrinsically correlated.\\
Ideally, spectroscopic data are needed not only to calculate redshifts,
but also to estimate the SFR and stellar mass properties \cite{Brinchmann2004,DelliVeneri2019}. However, spectroscopy is not always available and is extremely expensive in terms of observing time, becoming even more prohibitive when the goal is to characterize the properties of galaxies in large surveys. A potentially effective and alternative method of deriving such quantities in a combined framework, based on the exploitation of the supervised paradigm of machine learning, has recently been proposed. For example, Bonjean et al. \cite{Bonjean2019} approached a random forest model to simultaneously estimate SFR and stellar masses of galaxies from a sample of WISE NIR sources and related redshifts, by training the model on the SFR and stellar mass values extracted from spectra of the SDSS DR8. The encouraging results, although restricted to a limited redshift range, i.e. up to z less than $0.3$, prompted the use of this methodology to derive redshift and stellar mass in a combined way from a selected sample of galaxies from the DES survey, training the random forest on COSMOS2015 data, achieving interesting performances even in the regime of a limited space of photometric parameters \citep{mucesh2020}. The preliminary results demonstrate that ML could be a powerful methodology to find intrinsic correlations among photometric redshifts and galaxy physical parameters, which are extremely useful for large galaxy survey projects, like Euclid \cite{Laureijs2011,Bisigello2020}.\\

\subsection{Is the photo-z point estimate enough?}\label{sec:PDF}

\begin{figure}[!ht]
\begin{center}
\includegraphics[width=\textwidth]{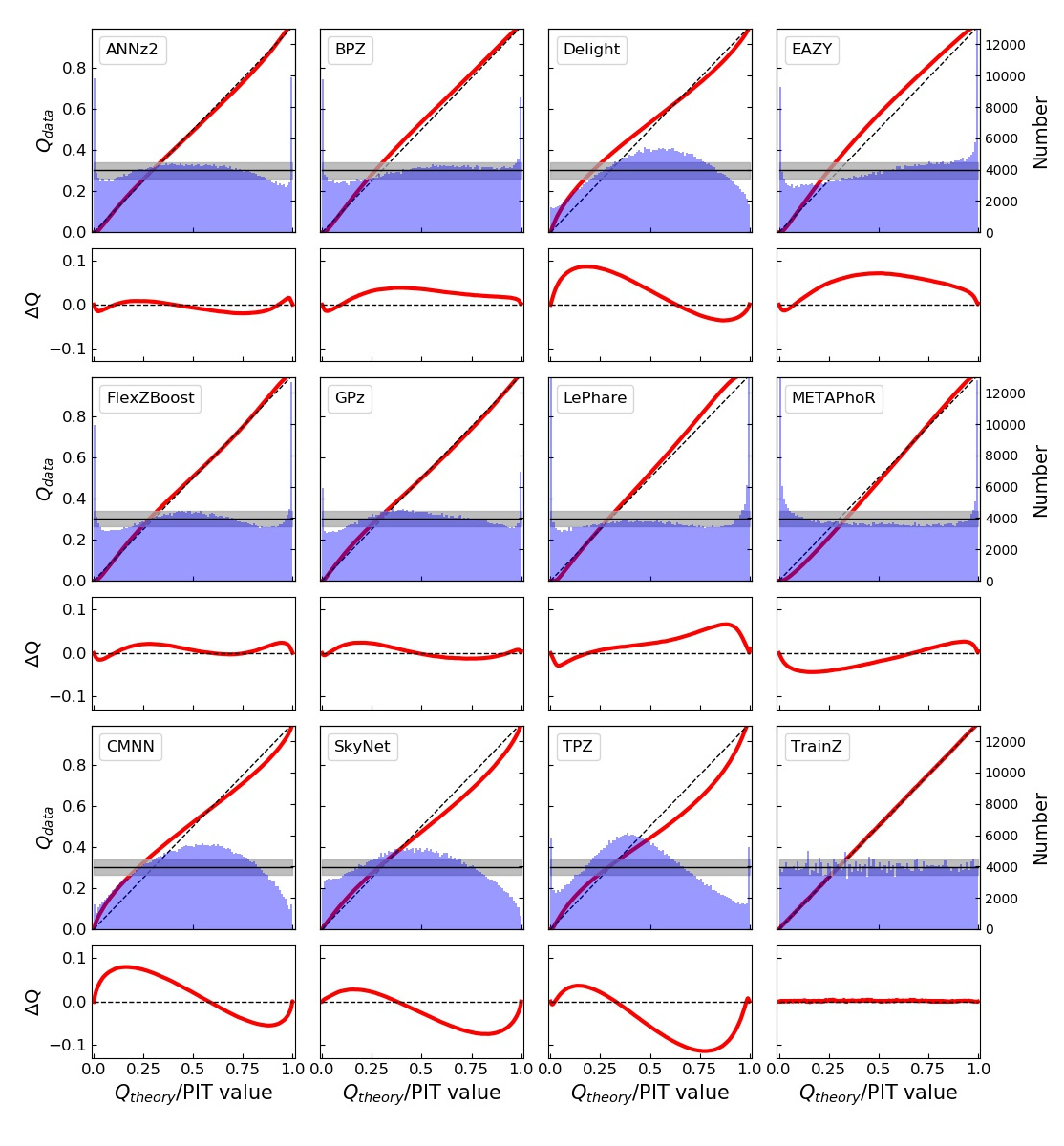}
\end{center}
\caption{Figure from \citet{Schmidt2020}. Statistical comparison among different methods for photo-z PDFs within the LSST Challenge (see \cite{Schmidt2020} for further details). The panels show the QQ plots (red) and PIT histograms (blue) with the ideal QQ (black dashed diagonal) and ideal PIT (gray horizontal) curves superimposed, including also a residual plot for the QQ estimation with respect to the ideal diagonal (lower inset). In the last bottom right panel TrainZ is the fake PDF described in Sec.~\ref{sec:PDF}.}\label{fig:lsst}
\end{figure}

Due to the wide variety of proposed solutions to the estimation of photometric redshifts, the need of a fair comparison among different methods is naturally needed. There are many studies in which people tried to compare their own method with what was available in literature \cite{Laurino2011,Brescia2013,Polsterer2016,DIsanto2018b,Brescia2019}. Furthermore, over the last decade, within large survey projects, there was the rise of challenges dedicated to the comparison among different methods following common rules and using same data, specifically provided for these contests. The PHAT (Photo-z accuracy testing) challenge \cite{Hildebrandt2010,Cavuoti2012} was among pioneers of such initiative, more recently followed by the Euclid and LSST challenges \cite{EuclidChallenge,Schmidt2020}. In all such contests a portion of the dataset, used for the final comparison analysis, was kept hidden to the participants in order to allow a blind test, thus providing a fair comparison in the same conditions (more examples are in  \cite{Abdalla2011,deJong2017,Bilicki2018,Tanaka2018,Amaro2019,Norris2019}), often with the purpose of testing a specific case of interest. The evaluation of the results was usually performed using standard statistical estimations, such as standard deviation, bias, normalized median absolute deviation, root mean square and outlier percentage rate, all metrics commonly adopted to evaluate the quality of photo-z predictions in terms of \textit{point estimates} and considered as sufficient and reliable for assessing the results.\\
However, in the last few years several studies showed that the use of the point estimation to fully represent the quality of photo-z's is insufficient and  could lead to biases \cite{Mandelbaum2008,Myers2009,Cunha2009,Wittman2009,Bordoloi2010,Abrahamse2011}. For such reason there was the tendency to adopt the Probability Density Functions (PDFs), in order to provide a wider confidence range on the photo-z prediction reliability, which could result particularly suitable in assessing the accuracy of photo-z estimation in cases where a higher precision is required, for instance to derive cosmological parameter measurements. For example, \citet{Mandelbaum2018}  demonstrated that the weak lensing studies, in particular the measurement of the critical mass surface density, require a reliable photo-z PDF estimation to remove any calibration bias effect.\\
Within few years, the trend to provide both photo-z point estimates and PDFs has now become a consolidated practice \cite{Sheldon2012,CarrascoKind2013,CarrascoKind2014,CarrascoKind2014b,Bonnett2015,Cavuoti2017b,Malz2018,Tanaka2018,Amaro2019,mucesh2020,Nishizawa2020}.
The idea is that a PDF should be able to provide a more complete information than the point estimation of the redshit. For instance, they should embed the presence of a secondary solution that, in presence of a degeneration of the parameter space, would be systematically suppressed. Some studies \citep{Viola2015,Mandelbaum2018} showed that the PDFs allow to improve the accuracy of cosmological and weak lensing measurements and to ensure a sufficient analysis of the cosmological uncertainties, from weak lensing tomography, to baryon acoustic oscillations. For this reason most of the surveys are now producing or are planning to provide photo-z catalogues including the PDFs and their statistics, rather than just the point estimates (see for instance KiDS \cite{deJong2017}, Euclid \cite{EuclidChallenge} and LSST \cite{Schmidt2020}).\\
For what concerns SED fitting methods it is well established the usage of the $\chi^2$ fit among data and a predefined set of galaxies \cite{Abdalla2011}, which leads directly to the derivation of a PDF by weighting all the possible solutions with their fit. Conversely, in the case of empirical methods, there is not such kind of homogeneity and different methods provide PDFs in different ways, spanning from the measure of the internal model error, by performing several independent training \cite{Sadeh2016}, to the measure of the effect due to the fluctuation in the parameter space (see \cite{Cavuoti2017b} and \cite{Schmidt2020} for a list of different methods with different strategies for PDFs derivation). It goes without saying that, in order to understand which is the best strategy (and the best method), in absence of an objective analytical tool, a comparison among different methods is required.\\
One of main differences among point estimates and PDFs is in the way in which they are evaluated and optimized. While for point estimates there is a common agreement on the statistical metrics, for what concerns PDFs, there is still no general agreement on how to assess their reliability. To demonstrate this, it is sufficient to refer to the examples of Euclid and LSST. In the first case, the quantities to optimize are the fractions of the stacked PDF enclosed in $\pm0.05$ or in $\pm0.15$ \cite{EuclidChallenge}, named $f_{0.05}$ and $f_{0.15}$ respectively. Such kind of metrics (as proved in \citet{Amaro2019}) can be easily falsified by using a simple \textit{dummy} PDF, consisting of a single bin PDF centered on the value of the photo-z point estimate.\\
\citet{Amaro2019} showed that, on a KiDS DR3 dataset with $z_{spec}$ less than one, a simple \textit{dummy} PDF is able to reach $93.1\%$ and $99.0\%$ in terms of $f_{0.05}$ and $f_{0.15}$, while well assessed methods, such as METAPHOR, ANNz2 and BPZ reach, respectively, $65.6\%$, $76.9\%$ and $46.9\%$ on $f_{0.05}$ and $91.0\%$, $97.7\%$ and $92.6\%$ on $f_{0.15}$, thus implying that those two parameters are only partially useful as metrics within an exhaustive evaluation process of photo-z PDFs.\\ 
In the second case (LSST, \cite{Schmidt2020}), main drivers are related to the property of the Cumulative Distribution Function (CDF), such as the Probability integral transform (PIT \cite{Seillier-Moiseiwitsch1993}) and the Quantile-Quantile plots (hereafter QQ, \cite{Wilk1968}). Within the same experiment those metrics were falsified through the usage of the same PDF for each point, corresponding to the redshift normalized distribution of the training set (see TrainZ panel of Fig.~\ref{fig:lsst}). While it is clear that the optimization of one of those estimators leads to meaningless PDFs, on the other hand the solution, that is to identify the correct estimator to optimize, is not at all clear and still remains an open issue.\\

\section{The critical role of the parameter space}\label{sec:PS}

The utility of photometric redshifts, derived from broad-band galaxy fluxes and colours rather than spectra, is now well established, through the high quality and reliability probed by many different techniques in a wide range of astrophysical contexts. The accuracy of the photometric redshift estimate certainly depends on the method used, but also on a complex combination of extension and distance of the galaxy, on the set of photometric bands, on the signal/noise ratio of photometry, as well as on the type of spectrum of the galaxy (in general, intrinsically redder objects produce more accurate photometric redshifts).\\

\subsection{The impact of the photometric uncertainties}
\begin{figure}[!ht]
\begin{center}
\includegraphics[width=10cm]{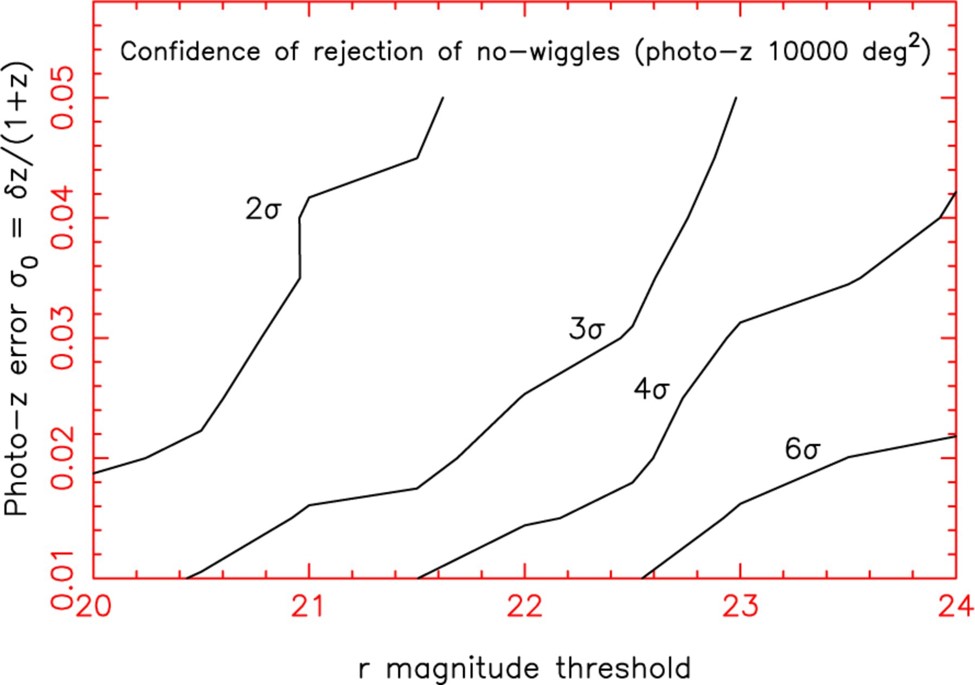}
\end{center}
\caption{Figure from \citet{Blake2005}. Confidence boundaries of photometric redshift imaging surveys with variable photometric error thresholds. The probabilities are expressed as a rejection ``$\sigma$ number'' for a Gaussian distribution. A detection area of $10,000 deg^2$ is used.}\label{fig:1}
\end{figure}

By considering the role played by the photo-z prediction error, the blurring of the large-scale structure in the radial direction, due to the photometric error of the redshift, degrades the measurements of the clustering pattern. Nonetheless, on physical scales greater than that implied by the redshift error, the information is preserved. On the other hand, even on smaller scales the large area covered by an image survey can potentially provide structural information very close to that produced by a fully spectroscopic survey, which implies very competitive cosmological constraints.
However, it is known that some sub-classes of galaxies exhibit better behavior in terms of photometric redshift. For example, experience with machine learning on SDSS data has shown that on a sample of luminous red galaxies we are able to obtain a redshift accuracy at least twice as high as that obtained on blue galaxies \citep{Brescia2014, Csabai2003, dabrusco2007}. 
In particular, the availability of sufficiently deep near infrared images is significant for galaxies with redshift $z > 0.4$ \citep{Bolzonella2000}. Therefore, the combined use of optical and infrared bands is able to improve the quality of distance predictions by exploiting the entire population of galaxies, rather than particular subclasses. However, Blake et al. \citep{Blake2005} have shown that a significant optical depth $(r\sim 24$) over an area of several thousand square degrees is required to ensure the accuracy of photometric redshifts useful for formulating measurements of cosmological properties (Fig.~\ref{fig:1}).

This photometric limit will be exceeded in the coming years, thanks to survey projects such as LSST, for which the photometric detection of the redshift will cover an A$\Omega$ area, approximating the entire sky to a depth of magnitude $r\sim 26$.\\
Modern precision cosmology requires very small statistical errors that, to be achieved, require the minimization of systematic errors through an in-depth knowledge of the various contributions to the loss of performance \citep{Oyaizu2008}. For example, it was estimated that for the tomographic investigations on a cosmic scale, based on the sampling of the dynamic range of distances, it is necessary to guarantee an uncertainty of  $\sim 0.003$ or less for the bias and dispersion in each bin of redshift, in order to control the constraints imposed by shot-noise on the estimation of dark energy \cite{Ma2006}. 
Furthermore, theoretically, a dependence on photometry errors would also be expected. The photometric simulations of the redshift and the subsequent modeling of the observed cases usually start from the assumption that the photometric errors follow a Gaussian distribution. However, the evidence of real data reveals a much more complex situation. Several redshift estimation experiments with machine learning methods have shown that the statistical quantities of the analysis of the prediction residuals, calculated with respect to the spectroscopic knowledge base, are always altered by anomalies induced by the tails in the distributions of photometric errors. The presence of a quantity of sources in the tails equal to about $10\%$ of the analyzed sample can cause an alteration in terms of dispersion of the prediction precision of more than $2\sigma$ compared to the impact of a halving of the S/N ratio \citep{Wittman2007}. This therefore implies the need to minimize the amount of error tails in magnitude and colour, especially for the photometric bands more sensitive to noise sources, such as the U band.\\
This problem obviously has also a strong impact in the tomographic analysis of the distribution of photometric redshifts, for which the only viaticum would be the substantial increase of the reference spectroscopic sample for the training of prediction models. Therefore, regardless of the photometric quality of the data, limiting the presence of tails in the photometric distributions can help to reduce the dependence on the spectroscopic sample, especially in the context of wide photometric surveys.\\
Concerning the contribution of the photometric errors to the accuracy of photo-z predictions with data-driven methods, despite the errors are not widely used in literature as input features to improve redshift estimation, there are few exceptions. For instance, in \cite{Laurino2011} the errors on each SDSS colour are used, while in \cite{DIsanto2018} the feature defined as $\sqrt{\sigma_{r_{model}}^2+\sigma_{r_{dev}}^2}$, where $\sigma_{r_{model}}$ and $\sigma_{r_{dev}}$ are the errors, respectively, on $model$ and $dev$ magnitudes in the r band\footnote{see: \url{https://www.sdss.org/dr12/algorithms/magnitudes} for further details.},  has been selected as the third more important for redshift estimation of QSO in the SDSS DR7.\\
Certainly, a positive aspect of data-driven learning methods is that they are automatically able to correctly learn the characteristics of the noise model. This requires that the learning data be characterized by a parameter space sufficiently extended to acquire the right information on the non-Gaussian characteristics of photometry, in addition to the need for the training and complete data sets to be homogeneous with each other in terms of uncertainty rate. In these cases, setting limits to the tails of the photometric error distribution can induce a greater control over the error model, minimizing the photometric dispersion of distances.\\

\subsection{The characterization of the photometric space}
\begin{figure}[!ht]
\begin{center}
\includegraphics[width=\textwidth]{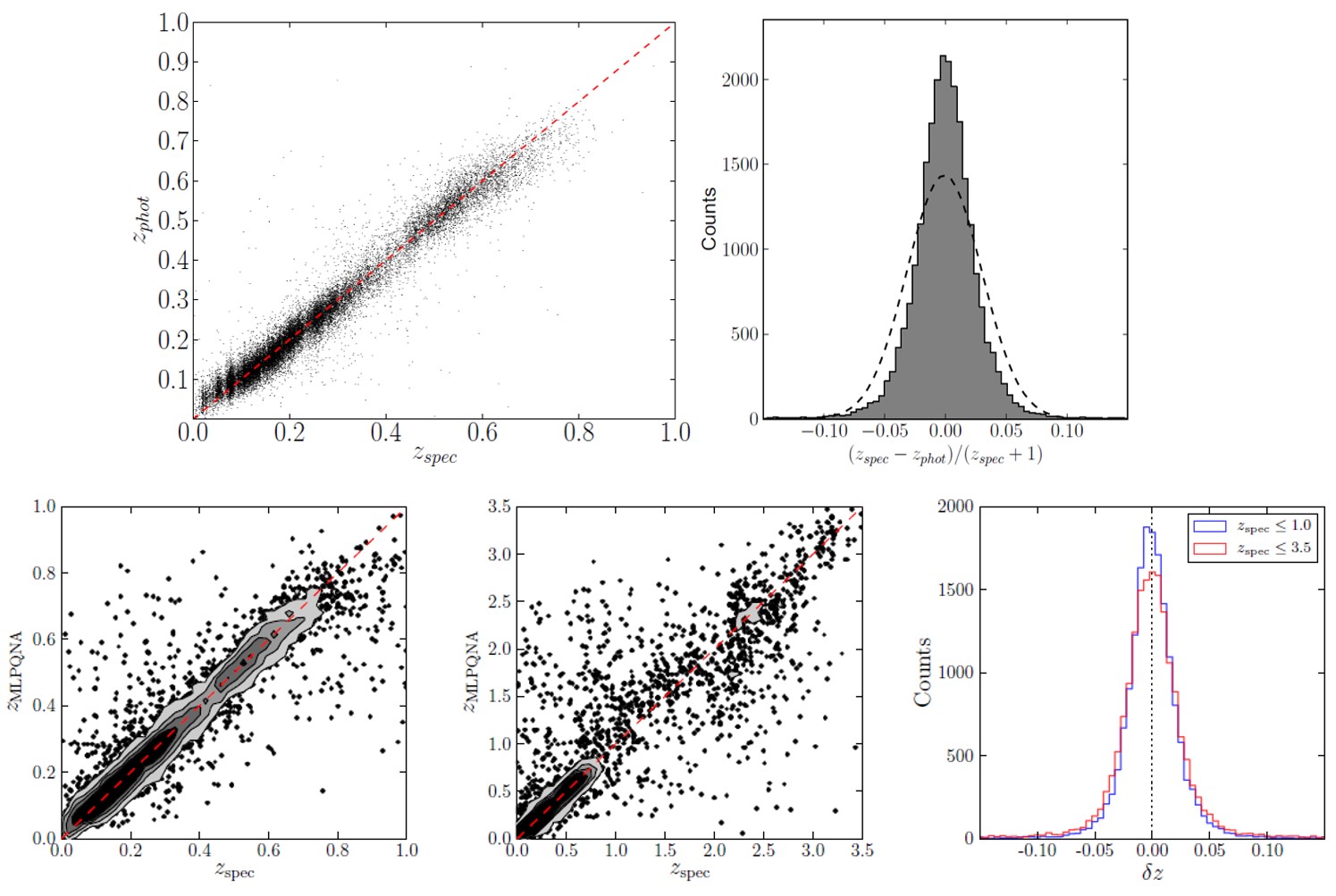}
\end{center}
\caption{Merged diagrams from \citet{Cavuoti2015} and \citet{deJong2017}. Upper panels: photo-z results obtained with KiDS DR2 data. In the right diagram the dashed line represents the Gaussian fit to the residual distribution. Lower panels: results of photo-z experiments with KiDS DR3 data. The left and central panes are the scatter plots for the experiments limiting the spectroscopic redshifts, respectively, to $1$ and $3.5$. On the right, the superimposed histograms of the residual distributions for the two cases are shown.}\label{fig:6}
\end{figure}

In the case of normal galaxies, it was experienced a high efficiency of photo-z estimation with the supervised machine learning model MLPQNA, a neural network based on the Multi-layer Perceptron with two hidden layers \citep{Rosenblatt1963}, which uses the Quasi Newton approximation of the Hessian error matrix as learning rule \citep{Nocedal2006}. This model performed particularly well for galaxies from the SDSS DR9 \cite{Brescia2014} and in the PHAT (Photo-z accuracy testing) challenge \cite{Hildebrandt2010,Cavuoti2012}, achieving excellent statistical results, according to the usual set of metrics, i.e. bias, scatter and outliers rate. This level of prediction accuracy was particularly unexpected in the case of the PHAT contest, which was a sort of worst case for machine learning methods. In fact, in that case the very limited amount of training data ($\sim500$ sources) evidenced their applicability limits with respect to SED fitting techniques, whenever the knowledge base is strongly lacking, regardless the intrinsic wavelength extension and quality of the photometry \cite{Hildebrandt2010}.\\ 
Another interesting use case was the photo-z estimation of galaxies provided by KiDS, a wide-area optical imaging survey in the four filters (u, g, r, i),
performed by the VLT Survey Telescope and the OmegaCAM camera \cite{deJong2013}. These redshifts are currently being used
by the KiDS collaboration for several studies related to the evolution
of galaxy stellar masses and the structural parameters with redshift \cite{Roy2018,Scognamiglio2020,Tortora2020}.\\
The KiDS DR2 contains $148$ tiles observed in all four filters \cite{deJong2015}. In order to derive the photometric redshifts, the multi-band source catalogues, based on source detection in the r-band images was used, for which it was extracted the training sample after having filtered objects having close and bright companions, affected by blending or bad pixels. The training spectroscopic redshifts were composed by merging data from SDSS DR9 and GAMA DR2 \cite{Driver2011}, therefore dominated by GAMA galaxies at low-z ($z \lesssim 0.4$), and by SDSS at the higher redshift regime (out to $z \sim 0.7$), with $r < 22$, while using $4$ and $6$ arcsec diameter apertures for the photometry. Thus obtaining an overall $1 \sigma$ uncertainty of $0.0305$ with a very small average bias of $0.0011$, a low NMAD of $0.021$, and a low fraction of outliers, i.e. $0.39\%$ above the standard limit of $0.15$ \cite{Cavuoti2015}.\\
Main differences between the KiDS-ESO DR3 and previous releases were the inclusion of \textit{GAaP} type magnitudes \cite{Kuijken2015} and the combined set of $440$ survey tiles, including large contiguous areas and achieving a refinement of the photometric calibration that benefits both the overlap among single filter observations and the stellar colours across filters. In the specific case of the KiDS-ESO DR3, two distinct experiments within different spectroscopic ranges were performed, respectively, $0.01 \leq zspec \leq 1$ and $0.01 \leq zspec \leq 3.5$. The statistics obtained in the first case were a bias = $0.0014$, $\sigma = 0.035$ and $NMAD = 0.018$, with $0.93\%$  of outliers ($\lvert \Delta z \rvert>0.15(z_{spec}+1)$); while in the second case a $bias = 0.0063$, $\sigma = 0.101$ and $NMAD = 0.022$, with $3.4\%$ of outliers rate were reached. These results are shown in Fig.~\ref{fig:6}. In terms of accuracy, within the spectroscopic limit of $zspec \leq 1$ our model (also in this case a MLPQNA, \cite{Cavuoti2015}) shows comparable results, while, as expected, the scatter and outlier rate efficiency decreases at fainter distances, due to the lower amount of sources available within the training sample. By looking at the histograms of the residual distributions of  Fig.~\ref{fig:6}, a peculiar behaviour appears, very frequent in the case of photo-z prediction with machine learning methods as also occurred for photo-z estimation in the SDSS galaxy experiments \cite{Brescia2014}, characterized by a \textit{leptokurtic} and symmetric distribution, i.e. an over-density of sources within the central region, populated by objects with a small error, which also reflects on the very low percentage of outlier rates and a low NMAD value.

\begin{figure}[!ht]
\begin{center}
\includegraphics[width=10cm]{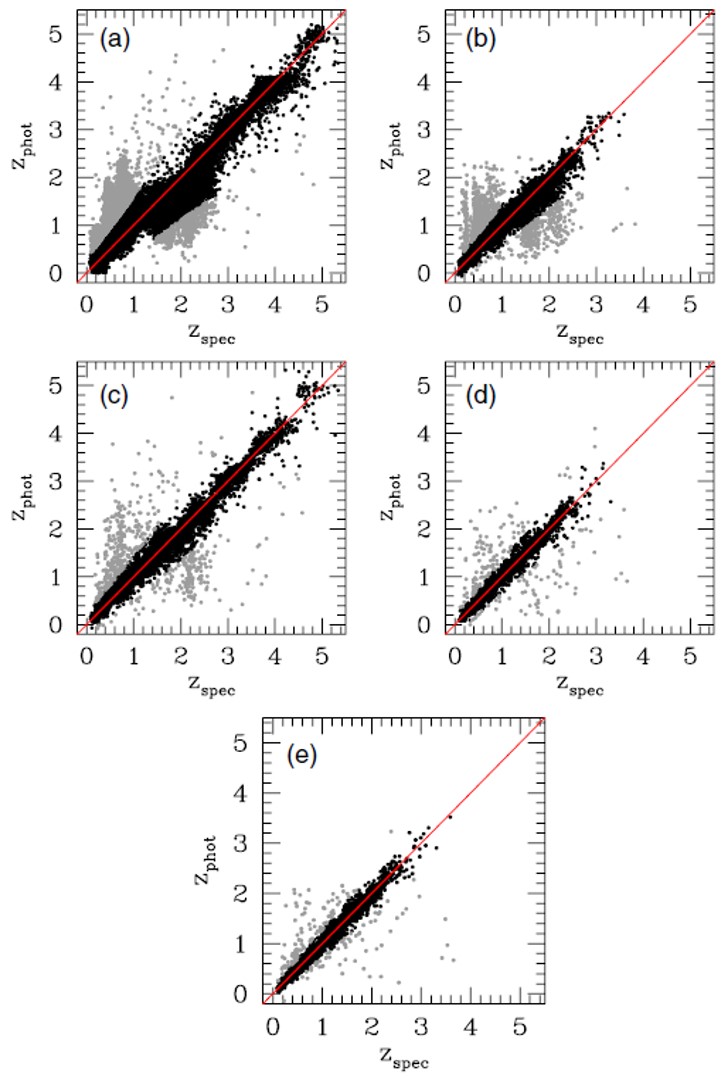}
\end{center}
\caption{Figure from \citet{Brescia2013}. Scatter plots of photo-z predictions for QSOs referred to the multi-band photometry using bands from, respectively, (a) SDSS, (b) GALEX+SDSS, (c) SDSS+UKIDSS, (d) GALEX+SDSS+UKIDSS, and (e) GALEX+SDSS+UKIDSS+WISE. Gray points are catastrophic outliers.}\label{fig:7}
\end{figure}

The MLPQNA neural network was also applied to the evaluation of photometric redshift for optically selected quasars \cite{Brescia2013} using a multi-wavelength photometric space composed by GALEX, SDSS, UKIDSS and WISE data, achieving very good levels of accuracy (bias = $0.004$ with a standard deviation of $\sigma = 0.069$) and a reduction of the number of catastrophic outliers to less than $3\%$. The comparison of performances reached by varying the number of bands (Fig.~\ref{fig:7}), from the optical SDSS photometry to the complete multi-band photometry from UV of GALEX to mid-IR of WISE, clearly shows that also in the case of QSOs, a wide photometric coverage improves the quality of predicted photo-z, also by using \textit{psf} type magnitudes, instead of the aperture type, more efficient for galaxies \cite{Brescia2013}.

\begin{figure}[!ht]
\begin{center}
\includegraphics[width=14cm]{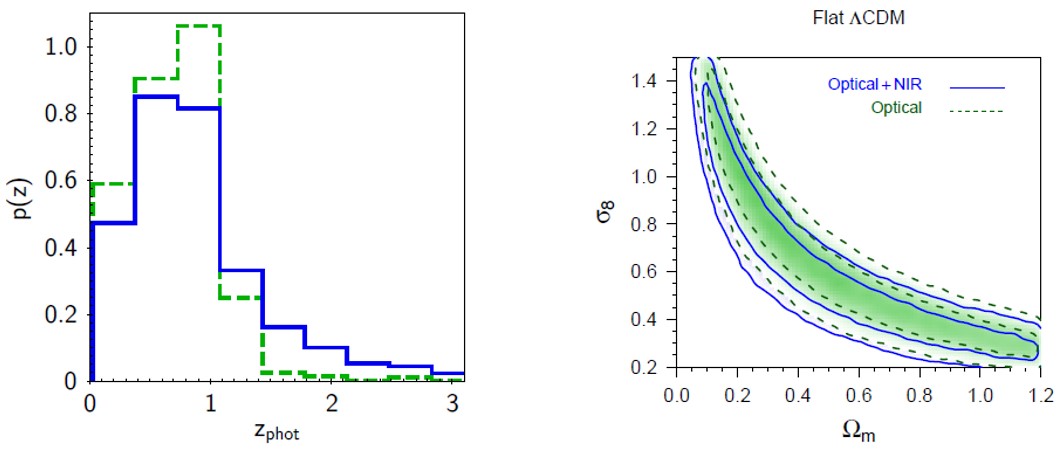}
\end{center}
\caption{Figure from \citet{Fu2018}. Left panel: The normalized histogram of photo-z estimated using photo-z derived using, respectively, four optical bands (green dash line) and eight (optical plus near-IR) bands (blue solid line). Right panel: Diagram of marginalized posterior density contours ($68.3\%$ and $95.5\%$) for $\Omega_M$ and $\sigma_{68}$ in the case of flat $\Lambda$CDM. The blue contours are the constraints related to the 8-band photo-z, while in green the results using 4-band photo-z are shown.}\label{fig:3}
\end{figure}

Another example of the positive contribution of near-IR bands to the photo-z prediction accuracy was discussed by \citet{Fu2018}, where they compared the cosmological constraints of $\sigma_{68}$ and $\Omega_M$, under the $\Lambda$CDM model, obtained by using the photometric redshifts derived from two different parameter spaces, respectively, with only the four optical bands ($u, g, r, i$ from VOICE \cite{Vaccari2016}) and with eight bands, by adding the near-IR bands ($Y, J, H, Ks$, obtained from VIDEO \cite{Jarvis2013}). The comparison, shown in Fig.~\ref{fig:3}, clearly reveals that in the 4-band photo-z case the contours appear shifted to the higher $\sigma_{68}$ and $\Omega_M$ side, coherent with the fact that the near-IR contribution correctly assigned the $15\%$ of the high-z galaxies to low-z regime (see \cite{Fu2018} for more details).\\
In general, therefore, given that the photo-z estimation is particularly crucial for acquiring knowledge about the formation and evolution of galaxies, by expanding the statistical sample available with respect to spectroscopic distances, it is possible to obtain a reliable knowledge of the distances even in regions of photometric space usually less covered by spectroscopy. Naturally, this implies a careful choice of photometric bands by large survey projects in the regions of interest, as well as ensuring the widest possible multi-wavelength coverage in order to minimize the occurrence of parameter space degeneration.\\

\subsection{Anomalies in the training set}
\begin{figure}[!ht]
\begin{center}
\includegraphics[width=12cm]{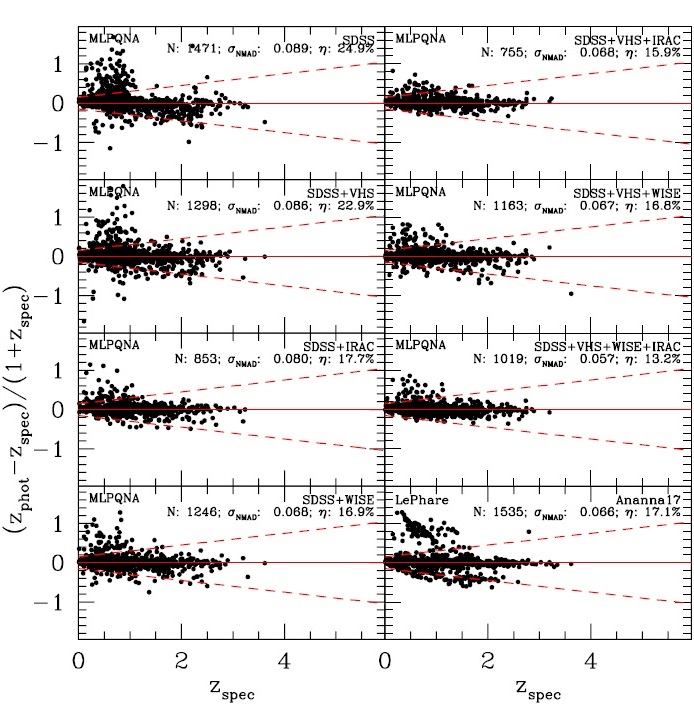}
\end{center}
\caption{Figure from \citet{Brescia2019}. Comparison between spectroscopic redshift and photo-z for the sources cut at the eROSITA flux and divided on the basis of
available photometry. For comparison, the result from \cite{Ananna2017} are reported in the lower right panel. By comparing the accuracy and the fraction of outliers in all panels, it appears clear that using only optical bands for bright X-ray sources is not sufficient.}\label{fig:4}
\end{figure}

\begin{figure}[!ht]
\begin{center}
\includegraphics[width=8cm]{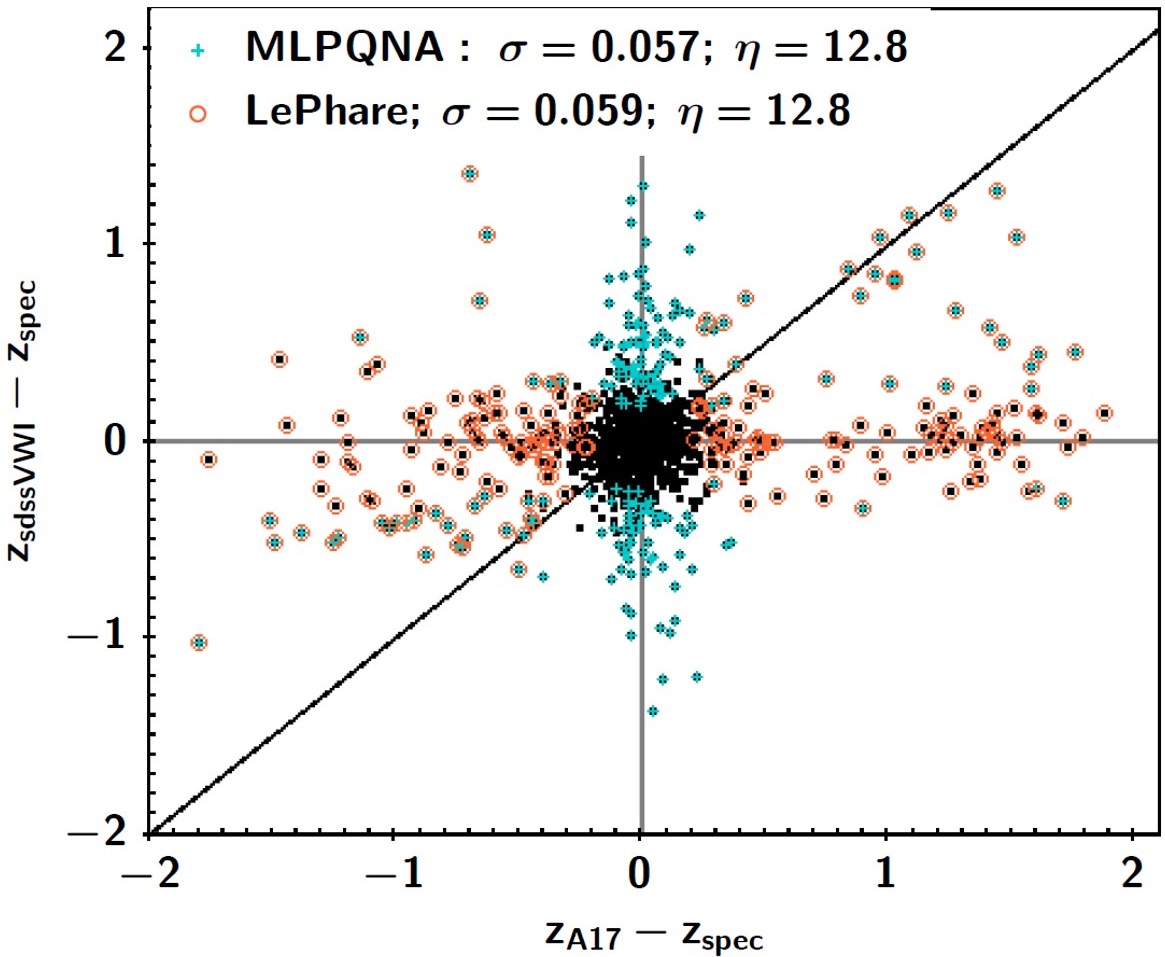}
\end{center}
\caption{Figure from \citet{Brescia2019}. Difference between spectroscopic redshift and photo-z computed via the MLPQNA neural network and the SED fitting method LePhare for the sub-sample including the complete photometry available (SDSS, VHS, WISE and IRAC), regardless their X-ray flux. Sources that are outliers for MLPQNA (LePhare) are plot in cyan (orange). For this sub-sample the two methods have a similar performance behaviour. Nevertheless, the majority of the outliers are such only for one of the two algorithms. For the common outliers along the  black one-to-one line, the two methods agree in terms of predicted photo-z.}\label{fig:5}
\end{figure}

Another important aspect in the photometric redshift prediction experiments concerns the identification of anomalies in the training set, potentially capable of leading to erroneous distance estimations. This aspect is directly related to the identification of regions of the photometric and spectroscopic parameters space that are under-sampled from the training data. This analysis is particularly important, as it can provide useful information to optimally and efficiently guide the follow-up spectroscopy, in order to optimize the training data set.
This information can also be used to evaluate the effectiveness of different combinations of photometric features by obtaining a statistical prediction of the redshift quality. For example, in the case of the Random Forest model, it is possible to verify the informative contribution of the photometric features, using the Out-Of-Bag (OOB) sampling technique, which consists in the random extraction of a sample of data excluded from the training during the construction of the various decision trees. This sub-sample of extracted data can then be used to estimate the relative importance of each photometric feature present in the parameter space. It is therefore an effective method to identify and remove features whose information entropy is redundant or even misleading \citep{Breiman2001}. \\
The informative contribution of the features can also be used to better understand the training data, to verify if it is possible to reduce the dimensionality of the problem and to identify areas of the space of the mapped parameters where the new training data can be incorporated in the most effective way. This prerogative can be obtained with supervised methods, as in the case of the TPZ model \citep{CarrascoKind2015}, and unsupervised, as in the case of the Self Organizing Map \citep{Masters2015} model. In the second case, the spectroscopic information on the target is not used in the process of constructing the Kohonen maps \citep{Kohonen1989}, but only offline, to identify the objects that belong to a cell in order to make predictions from the two-dimensional map.\\
In the case of Active Galactic Nuclei (AGN), the huge potential of the catalogues available for their science remains practically untapped, because most sources lack a redshift. The above techniques are routinely applied to galaxies, but their application to AGN (where the nuclear contribution to the global emission is unknown and depends on the type of source) is not straightforward. For this reason e.g. SDSS photometric redshifts have a low level of reliability for X-ray selected sources, especially at redshift below $z \sim 1$.\\ 
Nevertheless, recent studies have shown that the hybridization of empirical and SED fitting methods can provide encouraging results in the estimation of photo-z for mixed populations of galaxies and AGNs \cite{Duncan2018,Fotopoulou2018}, although the efficiency achieved for galaxies hosting AGN is not comparable to that obtained for inactive galaxies. In fact, the problem is that the extent of the AGN contribution to the total emission in the various bands is a priori unknown, causing uncertainty in determining a correct set of template models in the case of SED fitting \cite{Salvato11,Ananna2017}. This, except in the case of Seyfert galaxies (with low redshift and low luminosity), where the quality of the photo-z can reach that of normal galaxies, as long as a narrow/intermediate filter band photometry is available, like in the case of the COSMOS survey \cite{Salvato09}. Similarly, for empirical models based on the supervised paradigm, the limit is the availability of a sufficiently large and complete training spectroscopic sample \cite{Budavari01,Bovy12}. In this respect it must be pointed out that since most spectroscopic samples are usually extracted from optically selected galaxy catalogues, this unavoidably leads to an unbalanced distribution of AGNs (or any other peculiar objects), which are underrepresented. The  effects of this bias on future radio surveys (such as those to be performed with SKA) is clearly shown in \citet{Norris2019}.\\
However, empirical methods, which are implicitly less sensitive to differences in photometry, offer better performance, but show the need to identify the most suitable photometric parameter space. In this scenario, the approximately 3 million sources that eROSITA (Extended Roentgen Survey with an Imaging Telescope Array, \cite{Merloni12}) should observe, constitute the positive turning point.
A recent work \cite{Brescia2019}, investigated the contribution provided by the photometry, with an incremental number of bands, of the counterparts of the X-ray sources detected in the Stripe 82X \cite{LaMassa13,LaMassa13b,LaMassa16}, to the quality of photo-z of AGN sources estimated with machine learning methods.
The photometric catalogue included GALEX, SDSS, UKIRT, VHS, SPITZER / IRAC and WISE with sufficient depth to detect X-ray sources at least to the depth of eROSITA \cite{Ananna2017}. The results of the comparison between spectroscopic and photometric redshifts, obtained by our neural network for the sources in each wavelength sub-sample and by a SED fitting model are shown in Fig.~\ref{fig:4}.
Looking at the various diagrams, as it is reasonable to expect, the photometric coverage limited to the optical bands causes an excess of high redshift values for sources that actually have a low redshift. This effect can be reduced by adding the mid-IR bands of WISE, which show a better contribution than the near-IR bands of VHS. In particular, the addition of these bands allows a drastic removal of outliers, although obviously it reduces the sample of available sources by about $35\%$. A further improvement in statistical accuracy is achieved by adding the IRAC bands, a case in which the empirical model proves to be better than the SED fitting method, both in terms of a lower rate of outliers and an almost total absence of systematics.\\
However, at the depth of eROSITA, the two methods turn out to be comparable, especially in terms of percentages of outliers, as it can be seen from the Fig.~\ref{fig:5}. This diagram is particularly significant since it shows the recurrent phenomenon of a low rate of source outliers common to both prediction methods. This reveals the problem of the dependence of the outliers on the method used, which however allows to exclude the peculiar nature, from the astrophysical point of view, of the uncommon outlier objects.\\

As a concluding remark, in the presence of a conspicuous parameter space, for example including magnitudes, colours, ratios on tens of photometric bands, the removal of less important features is able to improve the quality of redshift prediction. The photo-z prediction accuracy through empirical methods has a complex dependence on the observed source types, the amount and quality of photometric bands, the spectroscopic quality and coverage of the photometric parameter space and the size of training set.\\

\subsection{The impact of spectroscopic reliability}
Despite the existing plethora of photometric redshift estimation solutions, so far none of the methods has been found to be able to achieve the accuracy of the spectroscopic redshifts measurement, which is of the order of $\sim 10^{-3}$ (e.g. \cite{LeFevre2005,Biviano2013,Rosati2014,Karman2015,Scodeggio2018,Hasinger2018,Angora2020}). While the best quality of photometric redshifts, obtained with broad-band photometry, can reach an error of $\sigma \sim 0.02$ \cite{Brescia2014,Salvato2019,EuclidChallenge,Schmidt2020}, that is an order of magnitude higher.\\
The accuracy of the photometric redshifts, obtained through supervised learning, has a natural dependence on the degree of completeness and quality of the spectroscopic catalogues used as ground truth. In fact, the incompleteness of the spectroscopic sample, although usually accentuated in the faint part of the photometric parameter space, can induce an altered selection effect that can be found throughout the parameter space. In addition, the residual error in estimating spectroscopic distances can affect the reliability of the metrics used for the validation of training by machine learning models, thus directly affecting the quality of the photo-z.
The typical quality of spectroscopic redshifts, between $95\%$ and $99\%$, implies that between the $1\%$ and $5\%$ of the training sample is unreliable. And the real problem is that it is not possible to establish a priori which samples are contaminated by such spectroscopic uncertainties, compared to those induced by photometry. A further complication is the impracticability of manual analysis methods, based on visual inspection, in the catalogues obtained from large survey projects. Hence the need to explore automatic mechanisms and procedures able of distinguishing the different sources of uncertainty in the data available.\\
In a recent study, \citet{Razim2021} proposed a method to identify the unreliable spectroscopic sample and consequently isolate the set of sources whose photometric parameter space is correctly mapped onto the spectroscopic sample. Obviously, with the ultimate goal of improving the quality of the photo-z estimation.
By exploiting the spectroscopy  from COSMOS and the Deep Imaging Multi-Object Spectrograph (DEIMOS, \cite{Hasinger2018}) catalogues, together with the 30-band photometry of the COSMOS2015 catalogue \cite{Laigle2016}, used to produce the photometric redshifts, the proposed method was based on two machine learning models, respectively, the Self Organizing Map (SOM, \cite{Kohonen1982}) for data analysis and cleaning, the MLPQNA for photo-z prediction, and the SED template fitting photo-z catalogue, described in \cite{Laigle2016}, as an additional testing and validation tool.\\
As known, a SOM produces a topographic map formed by a grid of neurons becoming selectively representatives of the various input patterns and changing their topological location during the course of the competitive learning. The final topology of the grid identifies more dense areas that result as overdensities (proto-clusters or cells) within the output space, corresponding to subsets of input patterns sharing some similarities in terms of internal features.
The unsupervised model SOM was introduced to identify and reject the unreliable spectroscopic redshifts, by introducing the coefficient $K_\mathrm{spec}$ to quantify the difference between a given galaxy spectrum and the average spectra of all galaxies located in the same SOM cell after training and then rejecting objects above a given threshold in $K_\mathrm{spec}$. Such a coefficient had the multiple effect to reduce the outlier rate of about the $88\%$ and the $1\sigma$ scatter by a factor of $\sim2$ within the photo-z predicted by MLPQNA and at the same time to reveal a high sensitivity to the physical variance of the galaxy population, thus becoming a reliable parameter to evaluate the correct mapping between spectroscopic distance and the photometric parameter space of the sources. Furthermore, using the DEIMOS spectroscopic catalogue as validation set, they exploited the so-called galaxy occupation map concept,  to verify that the sources of the validation sample and those of the knowledge base used by the MLPQNA model for the estimation of the photometric redshifts, occupied the same area of the SOM map, thus ensuring a correct correspondence between photometry and spectroscopic distances. Such a procedure reduced the outliers rate from $\sim11\%$ to $\sim2\%$.

\begin{figure}[!ht]
\begin{center}
    \includegraphics[width=0.9\textwidth]{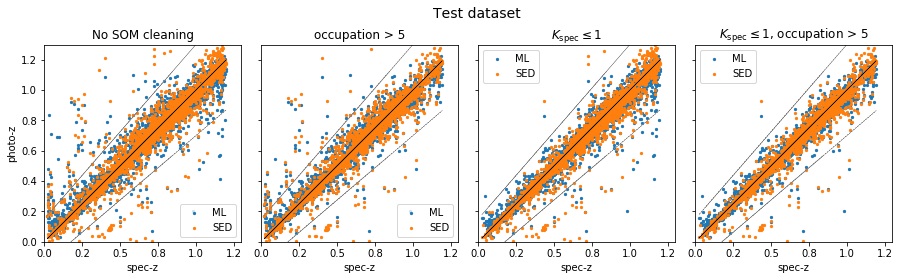}
     \includegraphics[width=0.9\textwidth]{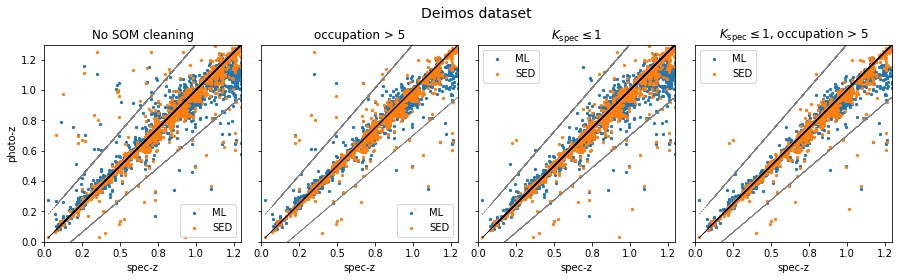}
     \includegraphics[width=0.9\textwidth]{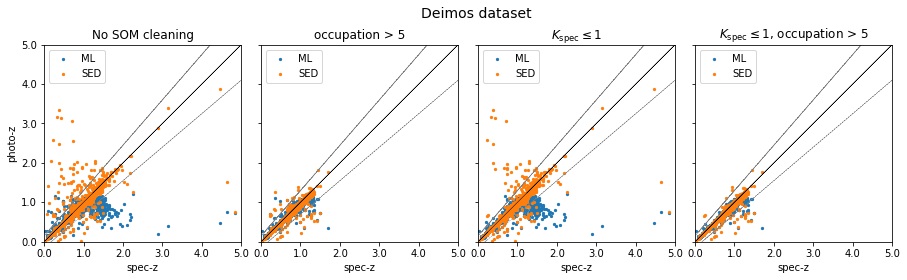}
\end{center}
\caption{Figure from \citet{Razim2021}. Scatter plots of machine learning and SED fitting photometric ($photo_z$) against spectroscopic ($spec_z$) redshifts. The photo-z were predicted by the MLPQNA neural network, trained on COSMOS spectroscopy and photometry and validated also on DEIMOS spectroscopic catalogue. The first left column of panels show the datasets before SOM filtering. The panels of the second column show the results after having filtered only the spec-z outliers. The third column panels report the plots after the filtering of only occupation maps, while in the last column the panels show the datasets after the combination of the two cleaning methods. The dotted lines show outlier boundaries defined as $photo_z = spec_z \pm 0.15$. The top first two rows show the results for the test and DEIMOS datasets limited to $spec_z<1.2$ and $photo_z<1.2$.}\label{fig:8}
\end{figure}

The scatter plots shown in Fig.~\ref{fig:8}, taken from \cite{Razim2021}, show the results of the proposed filtering method in terms of outliers reduction for the two involved machine learning and SED fitting models. In particular, the third row of panels demonstrates the predictions for the DEIMOS in the whole range of redshifts. Both models evidently show their need of the proposed occupation map filtering, in order to select the correct sources. Otherwise, as shown in the panels of the first column, SED fitting results affected by an important amount of catastrophic outliers in the whole range of spectroscopic redshifts, while the machine learning model tends to systematically fail for sources with $spec_z>1$. Furthermore, the filtering method appears particularly efficient in the case of outliers produced by overestimated photometric distances.\\
It goes without saying that a proper coverage of the parameter space is the only viable way in order to obtain reliable photo-z. In fact, in presence of portions of the parameter space not properly covered by the knowledge base, it would cause a proliferation of outliers. On the other hand, the presence of unreliable spectroscopic redshifts would result in a wrong training of the ML methods, inducing further bias hard to be handled.\\

\subsection{The importance of feature selection}\label{sec:FS}
\begin{figure}[!ht]
\begin{center}
\includegraphics[width=10cm]{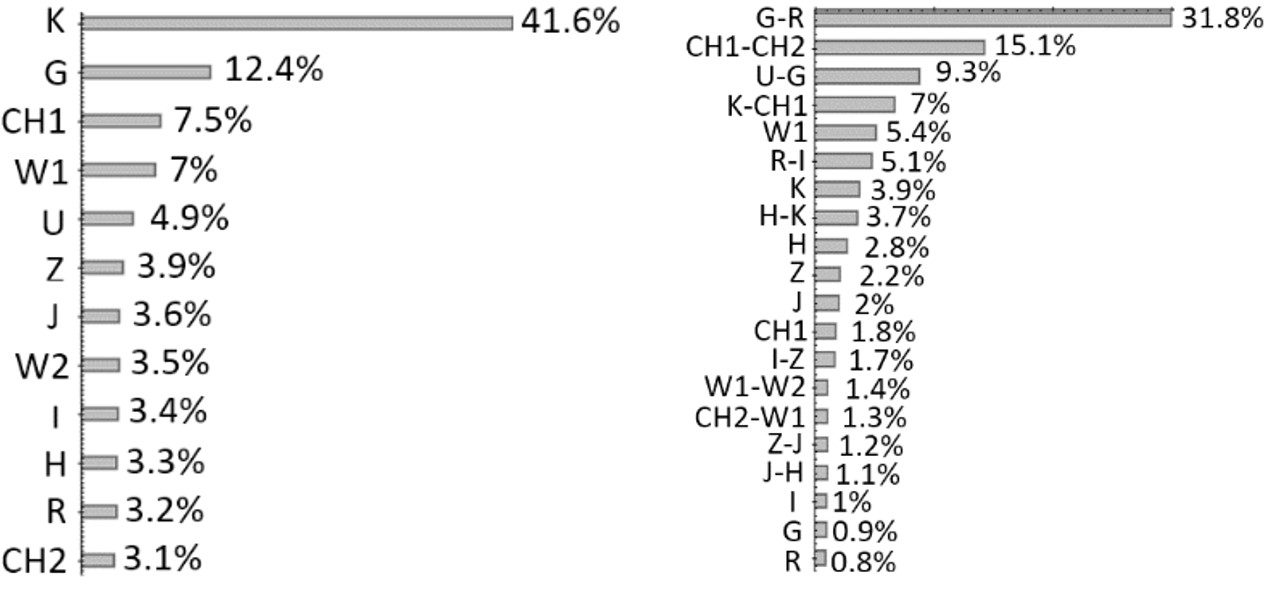}
\end{center}
\caption{Figure from \citet{Brescia2019}. Left: results of the feature analysis performed with $\Phi$LAB. The importance of each feature is estimated for the case in which only magnitudes are considered for the sample. Right: case in which a mixed parameter space (magnitudes and colours) is considered.}\label{fig:2}
\end{figure}

The concept of feature selection is linked to the property of the importance and relevance of features in the context of a parameter space used for prediction/classification purposes with methods based on machine learning. The importance of a feature is the relevance of its informative contribution to the solution of a learning problem, whereas a feature x is formally relevant if its removal from the parameter space always causes a degradation of the learning quality. Conversely, a feature x is considered weakly relevant if there is at least a subset A of features for which the accuracy of learning on A is worse than the union between A and x. In all other cases a feature is considered irrelevant.
Furthermore, on one hand, the computational cost of most machine learning methods scales badly with the number of dimensions. It proves therefore crucial to reduce the dimensionality by projecting the original space onto spaces of lower dimensionality; on the other hand, an important aspect of dimensionality reduction is to avoid overfitting if the 
number of dimensions is high.\\
This means that increasing information does not always correspond to an increase in knowledge in order to solve a problem, primarily due to the fact that the expansion of a parameter space inevitably causes an incremental dispersion of the correlation between the data, regardless from the metric used to define their mutual distances. A further crucial factor linked to feature selection is the possibility of obtaining a better physical interpretation of the phenomena underlying the problem addressed. In fact, by optimizing the parameter space, the features capable of solving/characterizing a problem are identified against those redundant or misleading.\\
These considerations introduce the taxonomy of approaches to feature selection: (i) \textit{most-relevant} feature selection, i.e. the selection of the smallest parameter space that provides the best accuracy. There are many methods proposed in the literature (cf. \cite{Guyon2003}), both for prediction and classification problems (Principal Component Analysis \cite{Jolliffe1986}, leave-one-out, forward selection, backward elimination, Random Forest \cite{Breiman2001}, PPS \cite{Staiano2005}, Naive-Bayes \cite{Ripley1996}); (ii) \textit{all-relevant} feature selection, i.e. the identification of the exact space of the parameters that are relevant to a variable extent for the solution of a given problem. Basically, in the second case a predictive/classification model is more likely to describe the various aspects of a problem, although it is necessary to increase the complexity of the feature selection method.\\
There are three general classes of feature selection methods. \textit{Filters}, based on arbitrary measures independent of any forecast/classification model and not designed to find complex correlations between features, thus unable to solve the all-relevant problem \cite{Gheyas2010}. \textit{Embedded}, which performs the feature selection at the same time of the prediction/classification model training execution, optimizing the feature set to improve accuracy. Such class is naturally designed to solve the most-relevant problem \cite{Guyon2003}. Finally, the \textit{Wrapper} class, a category in which the selection of features is performed by a dedicated prediction/classification model, in addition to the model used for the prediction or classification training task \cite{Kohavi1998}. As being specialized, it can use a deeper insight into the data than the filter class. Therefore, it can solve both most- and all-relevant problems.\\
One of the reasons why the all-relevant problem is more complex, such that only the methods of the wrapper class can address it, is that it is not always possible to use prediction/classification accuracy as a criterion for declaring a feature as not important. Indeed, the degradation of accuracy, upon removing a feature from the parameter space, is sufficient to declare the feature as important, but the lack of this effect is not sufficient to declare it as unimportant. In these cases, nothing can be said about the importance of one feature in combination with the others. Therefore, a more complex method for feature selection is required.\\
For the feature selection in the context of photometric redshifts prediction, early works relied on a trial-and-error approach. In other words, among the possible features, experiments were performed using all possible combinations of subsets of features, selected accordingly to the prescription of an expert. An alternative was the so-called \emph{data driven approach}, where a large subset (if not all) of all possible combinations of features are tried and the most performing one is selected, for instance by using a forward selection algorithm \citep{Guyon2003}, in order to identify the best set of feature for a given task (such an approach has been used with good results in \cite{Polsterer2016, DIsanto2018}). In \citet{Brescia2013}, the most significant features were selected by trying different combinations of magnitudes derived from a combination of surveys, respectively, GALEX (ultra-violet, \cite{Martin2005}), SDSS (optical), UKIDSS (near-IR, \cite{Lawrence2007}) and WISE (mid-IR, \cite{Wright2010}). 
This approach, however, adopted also in \citet{donalek2013} and \citet{DIsanto2016}, besides requiring a huge number of experiments and being therefore prone to computing limitations, does not ensure that the optimal performances are achieved. A more effective approach is to identify on objective grounds all the features which carry information useful to solve a given problem.\\
As known, Random Forest is one of the most suitable methods to perform the evaluation of the importance of features. It is mainly composed of a set (forest) of numerous simple predictors/classifiers (i.e. decision trees), each one built from different, randomly selected, combinations of feature subsets and data samples. During the learning phase, which corresponds to the forest tree building, each feature may have the same chance of being included in the decision chain, so even weakly relevant features will be statistically used in the forest construction process. The contribution of any feature can be easily calculated by considering all the trees that include that feature, so the contribution of both highly and weakly relevant features is well visible and measurable. Furthermore, Random Forest has a limited number of hyper-parameters and is relatively scalable with the data and parameter space sizes.\\
Recently, in \citet{Brescia2019} and \citet{DelliVeneri2019} it was introduced $\Phi$LAB (Parameter handling investigation LABoratory), a hybrid method, based on the exploitation of the Random Forest model, incorporating properties of both wrapper and embedding categories, thus designed to solve the all-relevant feature selection problem. The basic idea is the conjugation of two techniques, respectively, the inclusion within the parameter space of the so-called shadow features \cite{Kursa10}, a randomly noised version of real features and the $L_1$ norm regularization through Naive LASSO statistics (Least Absolute Shrinkage and Selection, \cite{tibshirani2013}). For instance, in \citet{Brescia2019} this method was evaluated by performing the feature selection on the multi-wavelength catalogue of the counterparts to the X-ray selected sources detected in Stripe 82X \cite{LaMassa16,Ananna2017}, to compare the quality of photometric redshifts estimations between our machine learning method MLPQNA \cite{Brescia2013} and the SED template fitting obtained by LePhare \cite{Arnouts1999,Ilbert06}. The optimization of the parameter space, composed in the specific case of magnitudes and colours, has allowed to extract a complete subset of high and weak relevant features capable of guaranteeing high precision in the estimation of redshifts and at the same time avoiding the degeneration of performance induced by the occurrence of redundant information. This together with the simplification of the problem by reducing the size of the space of the photometric parameters.\\
As it is shown in Fig.~\ref{fig:2}, the results of the all-relevant feature selection confirm an usual trend within the photometric redshift prediction cases, which has an intrinsic physical motivation. By considering a multi-wavelength parameter space composed only by  magnitudes, the K band is by far the most relevant feature. This can be easily motivated by considering that this rest-frame band corresponds to the knee of the galaxy SED, thus most suitable to determine the redshift than other bands. However, the relevance of this band and of magnitudes in general drastically changes by introducing the continuous information carried by colours, which become the most relevant features. In fact, by looking at the right panel of Fig.~\ref{fig:2}, the first four features collect more than $60\%$ of the total feature importance carried by the whole parameter space.

The reduction of the number of dimensions in a parameter space is not only convenient from the point of view of simplifying the computing complexity, but also and above all to overcome the well-known problem of the curse of dimensionality \citep{Bishop2006}, in which machine learning models exhibit a performance degradation when the number of features in the data representation space becomes significantly higher than the optimal one. \\

\subsection{Why not take advantage of astronomical images?}\label{sec:DL}
As evidenced from previous discussions, photometric redshifts are characterized by two main limitations. First, their precision decreases as the true redshift increases and, second, they are affected by the degeneracy between photometric colours and the spectroscopic redshifts, which means that within the source catalogues there is a plenty of objects sharing same colour band ranges but at different redshift.

For such objects any supervised machine learning method working with a photometric parameter space restricted to magnitudes and derived colours would be hopelessly induced into confusion. 
Therefore, as also stated by \citet{Hildebrandt2012}, one of the strongest limiting factor of both empirical and SED fitting techniques is the input information type used, i.e. the photometric measurements. They are directly affected by blending sources, variations of the Point Spread Function, being driven by the chosen aperture sizes or by the magnitude models, which are able to bring a limited fraction of the information that is potentially available from observed images.\\
The recent widespread diffusion of the deep learning paradigm \cite{Lecun2015} has also involved various fields of Astrophysics, from the morphological classification within a population of galaxies, to strong lensing, time domain astronomy and cluster membership recognition \cite{Dieleman2015,PasquetItam2018,Metcalf2019,Angora2020}.\\
In a deep learning model, both tasks of extraction of the input paramer space and the self-adaptive optimization are embedded into a single model. The first task is performed by the deep part of the model through a series of convolutions with specific filters and pooling operations. While the last smaller part of the hierarchical architecture is dedicated to the optimization task, which can be performed by any kind of traditional machine learning model. 
The outstanding property of deep learning to automatically extract features from images, like colour gradients, disk inclination, peculiar shapes, size and surface brightness of galaxies, opens a new and very promising perspective in the photo-z estimation field, becoming an efficient alternative to the manual feature selection, in particular by avoiding biases introduced during manual extraction and selection. There are already several works proposed in this respect, reaching high photo-z accuracy, at least competitive with other machine learning techniques based on boosted decision tree, feed-forward neural networks or random forest. 
Chong et al. \cite{Chong2019} proposed a Convolutional Neural Network (CNN) to predict galaxy morphological shapes, provided through Galaxy Zoo \cite{Willett2013}, to determine accurate photometric redshifts.
Hoyle \cite{Hoyle2016} exploited a CNN on multi-colour SDSS galaxy images, by splitting the spectroscopic redshift distribution into several bins, deriving a probability for any source to belong to those bins and assigning its redshift based on the most likely bin.
An hybrid deep/machine learning, based on the combined use of a CNN and a Mixture Density Network, was the choice of D’Isanto \& Polsterer \cite{DIsanto2018b} to obtain accurate photo-z from SDSS image cutouts and colours, derived by a pairwise subtraction of images, of galaxies and quasars.
The five band images from the flux-limited spectroscopic Main Galaxy Sample (MGS) of the SDSS were used as input data of a CNN by Pasquet et al. \cite{Pasquet2019} for photo-z estimation. We take this case to highlight an interesting property of deep learning applied to photometric redshift estimation. 

\begin{figure}[!ht]
\begin{center}
\includegraphics[width=12cm]{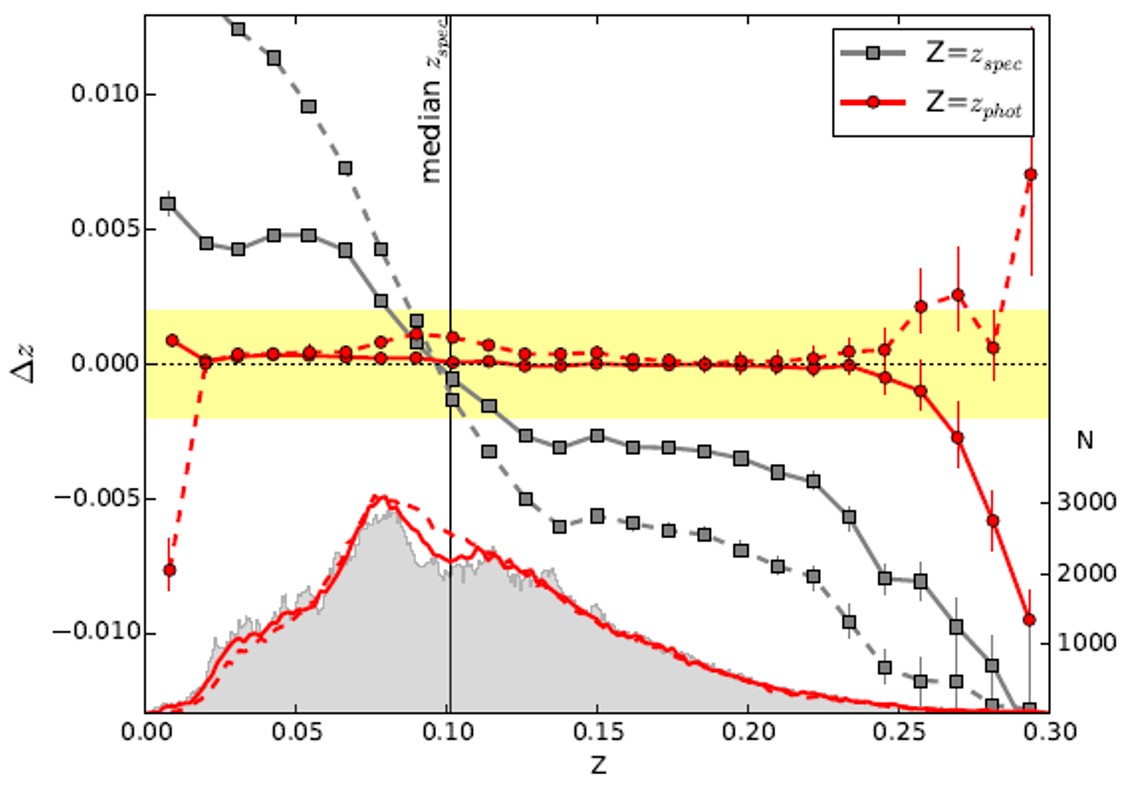}
\end{center}
\caption{Figure from \citet{Pasquet2019}. The bias as a function of both spectroscopic (gray) and photometric redshifts predicted by the CNN (red), with the corresponding redshift distributions and the comparison of results with \cite{Beck2016}. The yellow shaded zone represents the requirement for Euclid ($\Delta z \leq 0.002$) in photometric redshift bins.}\label{fig:9}
\end{figure}

As it is shown in Fig.~\ref{fig:9}, presented in \cite{Pasquet2019}, the CNN predicts photo-z estimated slightly outside the median redshift of the training sample. This implies the presence of a residual bias towards the most crowd redshift bins. However, such bias appears strongly limited in $1\sigma$, significantly smaller than the bias induced by the kNN method, used for direct comparison. In particular no any bias as a direct function of photo-z (such as galactic extinction or galaxy inclination) is found. \\
The application of deep learning is to be considered a prerogative extremely suitable for large survey projects, although still to be validated by comparing deep learning with more accurate machine learning models and by performing a double check between image and tabular features.\\

\section{Conclusions and perspectives}\label{sec:conclusions}
Astronomy is by definition a data-intensive science, especially by considering the incoming and future photometric survey projects, such as LSST, Euclid and JWST, all examples that will require data processing and storage solutions in peta- and exa-scale regimes. In such a context, data-driven approaches are not an option and a massive exploitation of deep learning paradigms seems to be the only chance to provide feasible solutions for analyzing those datasets. The massive exploitation of deep learning could open the possibility to predict photometric redshifts at the pixel level of calibrated images, instead of just using the limited and biased information carried by pre-processed catalogues. An indirect benefit of such strategy would also minimizing the serious problem of the right selection of the photometric space, having to choose among different apertures, psf or model magnitudes, luptitudes \cite{Lupton1999}, derived colours or magnitude ratios \cite{DIsanto2018}.\\
Using fully data-driven methods, such as unsupervised models, it is possible to identify regions of the multidimensional feature space in which every single method performs better. Thus providing important insights not only on the methods themselves, but also within the parameter space at different redshift regimes. Moreover, the data-driven paradigm can be successfully employed to verify the right coverage balance between photometric and spectroscopic spaces, to perform combined predictions of distances and galaxy physical parameters, as well as to disentangle different error contributions to the training data.\\
We are convinced that the present and future trend, driven by the demanding initiatives of large photometric surveys, is based on the photo-z challenges, in which several methods are carefully evaluated in a common and standardized framework, including same real/simulated training and blind testing data as well as metrics, to evaluate the strengths and weaknesses of each proposed solution. This is perfectly aligned with the recent satisfactory efforts to identify hybrid solutions, based on the combined use of empirical models, SED fitting methods and Bayesian statistics, showing that the best solution to optimize the quality of photo-z is to mediate the different prerogatives, in order to exploit at best the different useful sources of information.\\

\section*{Author Contributions}
MB, SC and GL oversaw the structuring and editing of the article. All Authors contributed to the bibliography search and selection. In particular, VA contributed to sections~\ref{sec:Plethora} and \ref{sec:PDF}. OR contributed to sections~\ref{sec:Intro} and \ref{sec:PS} and GR contributed to sections~\ref{sec:PS} and \ref{sec:FS}.\\


\section*{Funding}
MB acknowledges the funding from \textit{INAF PRIN-SKA 2017 program 1.05.01.88.04}. MB and GR acknowledge the financial contribution from the agreement \textit{ASI/INAF 2018-23-HH.0, Euclid ESA mission - Phase D}.
SC acknowledges the funding from \textit{Fondo di Finanziamento per le Attività Base di Ricerca (FFABR 2017)}.\\

\section*{Acknowledgments}
The Authors wish to warmly thank the colleagues Micol Bolzonella, Raffaele D'Abrusco, Antonio D'Isanto,  Amata Mercurio, Mara Salvato and Crescenzo Tortora for their contribution in many works and for their valuable suggestions in long and passionate discussions and exchanges of ideas. Furthermore, the Authors acknowledge the various contributions to their work offered by the survey projects CLASH-VLT, KiDS, Euclid and LSST and their Communities, in which most of the Authors are involved.


\bibliographystyle{frontiersinHLTH_FPHY} 
\bibliography{main}

\end{document}